\documentclass[12pt]{article}

\usepackage{epsfig}

\begin{document}




\centerline{\bf\large The dust origin of the Broad Line Region}
\vskip 0.5 true cm
\centerline{\bf\large  and the model consequences for AGN unification scheme\footnote{\rm Based on the talk presented during the COSPAR 2014 meeting; based on observations made with the Southern African Large
Telescope (SALT) under program 2012-1-POL-008 (PI: B. Czerny) \\ \vskip 1 true cm {\it Advances in Space Research \hfill 24 September 2014}} }

\vskip 2 true cm

\centerline{Bozena Czerny}
\centerline{\it Copernicus Astronomical Center, Bartycka 18, 00-716 Warsaw, Poland}


\vskip 0.5 true cm
\centerline{Justyna Modzelewska,  Francesco Petrogalli, Wojtek Pych,}
\centerline{Tek P. Adhikari,  and Piotr T. \. Zycki}
\centerline{\it Copernicus Astronomical Center, Bartycka 18, 00-716 Warsaw, Poland}
\vskip 0.5 true cm
\centerline{Krzysztof Hryniewicz}
\centerline{\it ISDC Data Centre for Astrophysics, Observatoire de Geneve, Universite de Geneve,} 
\centerline{\it Chemin d’Ecogia 16, 1290 Versoix, Switzerland}
\vskip 0.5 true cm
\centerline{Magdalena Krupa and Agnieszka Kurcz} 
\centerline{\it Astronomical Observatory of the Jagiellonian University, Orla 171, 30-244 Cracow, Poland}
\vskip 0.5 true cm
\centerline{Marek Niko\l ajuk}
\centerline{\it Faculty of Physics, University of Bialystok, Lipowa 41, 15-424 Bialystok, Poland}

\begin{abstract}

We propose a very simple physical mechanism responsible for the formation of the Low Ionization Line part 
of the Broad Line Region in Active Galactic Nuclei.  It explains the scaling of the Broad Line Region size with the monochromatic luminosity, including the exact slope and the proportionality constant, seen in the reverberation studies of nearby sources. The scaling is independent from the mass and accretion rate of an active nucleus. The mechanism predicts the formation of a dust-driven wind in the disk region where the local effective temperature of a non-illuminated accretion disk drops below 1000 K and allows for dust formation. We explore now the predictive power of the model with the aim to differentiate between this model and the previously proposed mechanisms of the formation of the Broad Line Region. We discuss the expected departures from the universal scaling at long wavelength, and the role of the inclination angle of the accretion disk in the source. We compare the expected line profiles with Mg II line profiles in the quasars observed by us with the SALT telescope. We also discuss the tests based on the presence or absence of the broad emission lines in low luminosity active galaxies. Finally, we discuss the future tests of the model to be done with expected ground-based observeations and satellite missions.


\end{abstract}

\parindent=0.5 cm

\section{Introduction}
\label{sect:intro}

Broad emission lines seen in the optical and UV spectra are the most characteristic features of active galaxies. They led to the discovery of quasars (Schmidt 1963), and they remain one of the most powerful tools to study the accretion processes onto supermassive black holes, the outflows from the nuclear region, and the role of active galactic nuclei (AGN) in shaping the host galaxy. Since the Broad Line Region (BLR) is very compact it remains unresolved. Our knowledge of this region relies on the spectral analysis, including the spectra variability, and on the theoretical interpretation of the processes responsible for the observed emission of radiation.

BLR properties have been studied in a numerous papers and books (for general reviews, see Krolik 1999, Netzer 2013). The structure of the region is complex, since part of the radiation comes likely from the surface of an accretion disk surrounding the central black hole and the other part comes from the gas well above the disk. The dynamics is equally complex, and the reverberation studies show both the Keplerian component to the cloud motion, further complicated by the possible spiral structures in the disk  as well as strong turbulent motion, inflow and outflow (e.g. Shapovalova et al. 2010, Grier et al. 2013).

It was postulated long time ago that the broad emission lines should actually be divided into two distinct components. Low Ionization Lines (LIL), like H$\beta$, MgII, or Fe II should form in higher density region, likely closer to the disk or within the disk. High Ionization Lines (HIL) are emitted by the lower density regions (Collin-Souffrin et al. 1988). The dynamics of the two regions seems also different: LIL does not show a strong inflow/outflow while an outflow is characteristic for HIL. The location of HIL is also close to the central black hole, as indicated by the time delays between the lines and the continuum.

In this paper we further concentrate on the LIL part. Their characteristic properties can be summarized in the following way: (i) line profiles is usually inconsistent with their origin in a Keplerian disk; in Narrow Line Seyfert 1 galaxies and type A quasars they are well represented by a simple Lorentzian symmetric profile; double-peak profile is seen only in a small fraction of objects (e.g. Eracleous et al. 2009) (ii) the lines do not indicate strong net inflow/outflow (iii) the covering factor is high; about 30 \% of the nuclear emission is intercepted by the corresponding part of the BLR (iv) the size of this region scales well with the square root of the monochromatic flux of an active galaxy, as showed in numerous reverberation studies (e.g. Kaspi et al. 2000; Peterson et al. 2004; Bentz et al. 2009).

This means that the material cannot be entirely close to the disk, since in that case the covering factor would be small. It cannot also be a massive wind or an inflow, since then the lines would be Doppler-shifted with respect to the host galaxy. However, the material has to have an opportunity to appear high above the disk to intercept enough of the central radiation flux.

So far three mechanisms of the formation of the BLR has been proposed in the past: (i) magnetic wind (Elitzur \& Shlossmann 2006; Elitzur \& Ho
2009; Elitzur et al. 2014), (ii) accretion disk instability due to self-gravity (Collin \& Zahn 1999; see also Collin \& Zahn 2008; Wang et al. 2012)  (iii) the outflow connected with the transition between the radiation-pressure dominance and the gas-pressure dominance in the accretion disk (Risaliti \& Elvis 2010).

Here we consider the fourth model: the Failed Radiatively Accelerated Dust-driven Outflow (hereafter FRADO) which well explains the above properties, and we analyze the predictions of this model which can be used to prove or falsify the proposed mechanism.

\section{Failed Radiatively Accelerated Dust-driven Outflow (FRADO) model of the BLR}

The model has been proposed by Czerny \& Hryniewicz (2011) and tested against the data for the most studied Seyfert 1 galaxy NGC 5548 by Galianni \& Horne (2013). Here we summarize the basic outline of the model.

The presence of dust was extensively discussed in the context of the dusty-molecular torus, which was introduced by Antonucci \& Miller (1985) in order to explain the spectra of Seyfert 2 galaxies. The torus is located much further from the nucleus than the BLR. At the distance of the torus the irradiation of the matter by the X-ray emission from the central corona and by the UV emission from the central parts of the accretion disk is not strong enough to destroy the dust particles. This is supported by the studies of the reverberation in the IR (e.g. Pozo Nunez et al. 2014). The torus also provides the outer radius to the BLR, as shown by Netzer \& Laor (1993). Closer in, the radiation destroys the unshielded dust. However, if the dust is not exposed to the irradiation, it may be present closer in, and the presence of the dust in the BLR was hinted at by several authors (e.g. Dong et al. 2008). Czerny \& Hryniewicz (2011) proposed a particularly simple and appealing picture with profound and specific consequences.

\begin{figure}

\label{fig:view}

\begin{center}

\includegraphics*[width=10cm,angle=0]{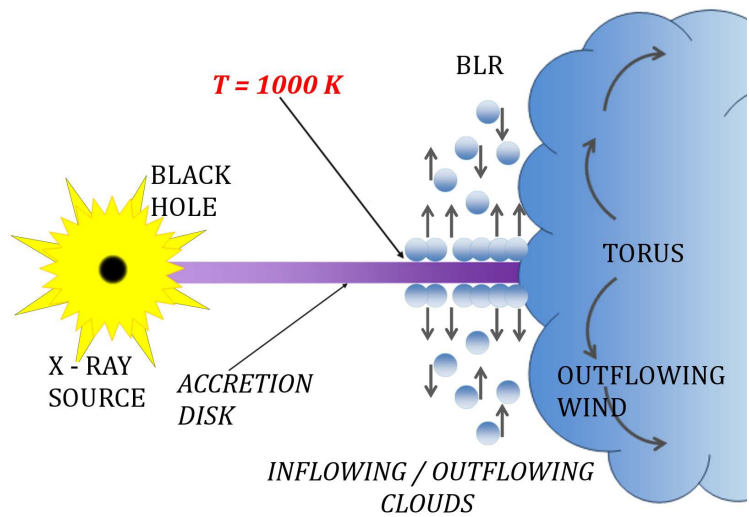}

\end{center}

\caption{This figure illustrates the basic mechanism of the formation of the BLR. Accretion disk is not strongly irradiated, and in the region with the disk effective temperature below 1000 K a dusty outflow forms. The matter rises high above the disk surface, irradiation increases, the dust evaporates, the radiation pressure deccreases and the material falls back onto the disk surface affected by gravity.}

\end{figure}

Since the accretion disks in AGN are not predicted to be very geometrically thick, the effects of the direct irradiation there are not as strong as in the case of binary systems (see e.g. Loska et al. 2004). The disk effective temperature, $T_{eff}$, at a given radius, $r$, is thus basically determined by the black hole mass and accretion rate. This temperature, of order of $10^5$ K in the innermost part of a quasar, decreases outwards as $r^{-3/4}$, for a stationary disk, according to the classical study of Shakura \& Sunyaev (1973)
\begin{equation}
T_{eff} \propto \left({ M \dot M \over r^3}\right)^{1/4}
\label{eq:teff}
\end{equation}  
where $M$ is the black hole mass and $\dot M$ is the accretion rate. When the temperature drops below the dust sublimation temperature of about 1000 K, the dust forms in the disk atmosphere. It is known from the studies of cold stars that dust easily leads to powerful dust-driven winds. Thus the radiation pressure rises the material high above the disk. There, however, the material becomes irradiated by the central source, the dust evaporates, the radiation pressure force decreases and the gravity pulls the material back towards the disk. This leads to a formation of a failed wind which does not escape the area. The kinematics of the matter motion consists of the original Keplerian motion imposed by the disk and the vertical positive and negative velocity of rising and falling clumps. The schematic view is shown in Fig.~1. The radius, where it happens, depends on the product of the two parameters: $M$ and $\dot M$ (see Eq.~\ref{eq:teff}).

The same theory (Shakura \& Sunyaev 1973) tells us, that there is another quantity, which also depends only on the product of   $M$ and $\dot M$. This is the monochromatic flux at a fixed frequency, $L_{\nu}$. It is a textbook result that the dependence on the frequency is   $L_{\nu} \propto \nu^{1/3}$ but the theory predicts also the proportionality coefficient that includes the product $(M \dot M)^{2/3}$ and universal atomic constants (e.g. Tripp et al. 1994)

\begin{equation}
L_{\nu} \propto \nu^{1/3} (M \dot M)^{2/3}.
\label{eq:lnu}
\end{equation}

Assuming that $T_{eff} = 1000$ K at the inner radius of the BLR, $R_{\rm BLR}$, we combine Eq.~\ref{eq:teff} and Eq.~\ref{eq:lnu} to obtain a universal relation 
\begin{equation}
R_{\rm BLR} \propto L_{\nu}^{1/2}
\end{equation}
for a continuum monochromatic flux measured at a fixed wavelength. It is now independent from the black hole mass or accretion rate. Calculating the coefficient properly, Czerny \& Hryniewicz (2011) obtained
\begin{equation}
\log R_{\rm BLR} = 1.541 + 0.5 \log L_{44,5100}{\rm~~~~  [light ~ days]}
\label{eq:RBLR1}
\end{equation} 
where $L_{44,5100}$ is the monochromatic flux measured at 5100 \AA~ in units of $10^{44}$ erg s$^{-1}$, and thus with high accuracy recovered the observational formula of Bentz et al. (2009). The constant present in the formula depends on the frequency used for the continuum measurement (here 5100 \AA), the dust sublimation temperature (here 1000 K), and the inclination angle (here $i = 39.2^{\circ}$, after Lawrence \& Elvis 2010). This last issue will be discussed below.

\section{Tests of the FRADO model}

Since there are other models of the formation of the BLR regions, each of them should be tested against observations. Here we treat the theory as a starting point, and formulate specific predictions which results from the assumptions underlying the model.

\subsection{Deviation of the disk continuum from a power law}

Eq.~\ref{eq:lnu} applies only at the long wavelength limit of an accretion disk spectrum, i.e. when we measure the continuum coming from the regions far enough from the disk interior where the disk temperature has its maximum. At shorter wavelengths the spectrum bends and finally, at the shortest wavelengths, it can be roughly approximated by an exponential cut-off, corresponding to the maximum temperature.

\begin{figure}
\label{fig:pldev}
\begin{center}
\includegraphics*[width=10cm,angle=0]{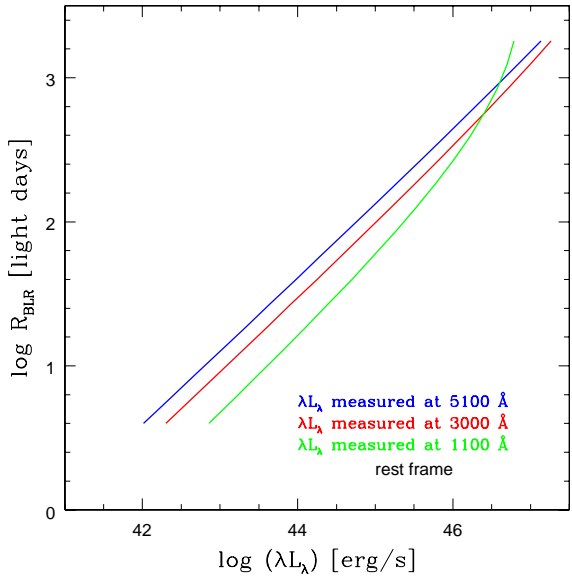}
\end{center}
\caption{The theoretical relation between the  BLR inner radius and the monochromatic flux measured at 5100 \AA, 3000 \AA~ and 1000 \AA, for the Eddington ratio of  0.3. Significant deviations from the universal law are seen only for the shortest wavelength. }
\end{figure}

In order to estimate the likely departures from a power law, we thus use the Novikov-Thorne model of an accretion disk (Novikov \& Thorne 1973, with later developments, see Czerny et  al. 2011 for the description). The model includes all relativistic effects but does not include advection (Abramowicz et al. 1988; S\c adowski et al. 2011) so it applies to the Eddington ratio not much higher than 0.3. We fixed the inclination angle at 30 deg, and we calculated a series of models for a range of black hole masses and for two Eddington ratios: 0.1 and 0.3. Higher Eddington ratio would require including the advection term into the model. The radius of the BLR was found from the disk model (where the $T_{eff}$ dropped below 1000 K) and the continuum was determined from the spectra at three frequently used values of the wavelength. The results for the spin black hole fixed at 0 are shown in Fig.~2.

We see that the measurement of the continuum at 5100 \AA~ produces a power law relation, the deviations for 3000 \AA ~ are also not very strong, so using this wavelength for high redshift quasars is still reasonable. However, going further towards shorter wavelengths gives significant deviations from the power law. The curve in Fig.~2 steepens for the highest values of the luminosity, and the size of the BLR is larger than expected from the simple scaling with monochromatic luminosity. This may cause problems for quasars with redshifts $z$ higher than 2 and studied in the optical band. In this case also the departures from the simple disk models also become significant, since the far-UV and soft X-ray band spectrum is strongly affected by Comptonization of the disk photons (e.g. Czerny et al. 2003, Davis et al. 2007). Only rare quasars do not show this problem (e.g. Hryniewicz   et al. 2010, Czerny et al. 2011).

\subsection{Dependence on the inclination angle of the accretion disk}

The emission of the accretion disk is optically thick and it is strongly anisotropic, with flux decreasing roughly with the $\cos i$, where $i$ is the inclination of the system measured from the symmetry axis. The relativistic effects slightly enhance this effect, but not strongly. Therefore, Eq.~\ref{eq:lnu} contains this factor (see e.g. the book by Frank, King \& Raine 2002). Also, if the system is inclined, the delay measured between the continuum and the BLR depends on the inclination. In general, since the region is extended, this delay should be described by a complex transfer function (see e.g. Grier et al. 2013). However, here for the purpose of a simple estimate of this effect, we describe the delay as in Czerny \& Hryniewicz (2011). We assume, that all the reprocessing happens in the equatorial region, located in the opposite side to an inclined observer. Thus the relation between the size of the BLR as measured from the time delay between the line and continuum (i.e. the monochromatic flux) should contain a factor
\begin{equation}
R_{\rm BLR} \propto L_{\nu}^{1/2} \cos i^{1/2} (1 + \sin i)
\label{eq:rcosi}
\end{equation}

This means that the relation size - luminosity should be broadened by the lack of knowledge of the inclination. If the sample of objects covers the range of the inclinations between 0 and 45 deg, randomly distributed in $\cos i$, then the mean inclination would be 31.4$^{\circ}$, and the largest/smallest values would deviate from the mean value in logarithm of the radius by $\sim 0.1$. This effect is not very strong. Still, we have made an experiment to see if including the knowledge of the inclinations helps to decrease the dispersion in the size - luminosity relation.

\begin{figure}
\label{fig:niko}
\begin{center}
\includegraphics*[width=10cm,angle=0]{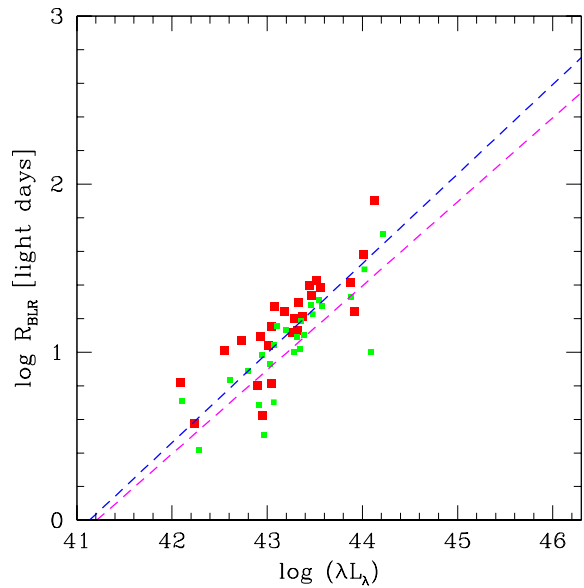}
\end{center}
\caption{The observed relation between the  BLR inner radius and the monochromatic flux from a subsample of Bentz et al. (2009) with inclinations measured by Niko\l ajuk et al. (2006) (red points, blue line), and the relation corrected for the inclination effect described in Eq.~\ref{eq:rcosi} (green points, magenta line). }
\end{figure}

For that purpose we took the sample of AGN from Bentz et al. (2009) and we selected sources with the inclination measured by Niko\l ajuk et al. (2006) using the X-ray variability (i.e. X-ray dispersion method). There are only 9 of such sources, and 24 measurements. The number of measurements is higher since the reverberation of some sources (e.g. NGC 5548) were performed during several periods long enough that they could be treated separately. The results are shown in Fig.~3. When we fitted the slope and dispersion in the selected subsample without the inclination effect, the dispersion was 0.139. However, when we corrected the delay and the luminosity for the inclination, dispersion was slightly reduced dawn to 0.128. We also tried to use the subsample with inclination angles determined by Nandra et al. (1997) from the fits to the X-ray Fe K$\alpha$ line. However, in that case the results were much worse, the dispersion was actually increased from 0.14 to 0.17. It is likely that the inclinations from the Fe K$\alpha$ in that sample were not yet measured reliably due to contamination by the narrow K$\alpha$ component. 

It is also possible that the disk inclination measured close to the black hole (from the shape of the iron K$\alpha$ line) is different than the inclination in the
distant BLR zone. The possible effects of the disk warps has been discussed in a number of papers  (e.g. Herrnstein et al. 1996 for NGC 4258; 
Caproni et al. 2006, Ivanov \& Papaloizou 2008, and the references therein for a general case) but the direct observational evidence exists only for maser sources like NGC 4258. Certainly, more AGN with the measured inclination are needed to perform further tests.

\subsection{The Mg II line shape}

The FRADO model can be used to predict the velocity field of the emitting plasma, which can be later converted into the line profile. Here we develop a very crude approximation. We consider the motion of a single dust grain as representative for the dusty cloud but we allow for the mass to be higher than just the dust grain mass to mimic the coupling between the dust and the gas. We neglect the irradiation from the central parts of an AGN,  we assume the effective cross section for the dust particle instead of a wavelength-dependent cross section for the radiation absorption, and finally we assume that the  departure from the disk plane is not very large. These approximations allow to obtain a very simple analytic solution to the motion in the vertical direction.

The equation of motion of the dusty particle in the vertical direction reduces to:

\begin{equation}
m{dv_z \over dz} = -{GMmz \over r^3} + {F \sigma_{dust}\over c}
\label{eq:dynamics}
\end{equation}
where $z$ is the vertical coordinate, $v_z$ is the vertical velocity, $m$ is the mass of the particle, $F$ is flux from the unit surface of the accretion disk at a given radius $r$, $\sigma_{dust}$ is the total (i.e. frequency-averaged) cross section for dust absorption and scattering, and $c$ is the light speed. The first term gives the gravitational attraction and the second therm is the radiation pressure force.

The flux $F$ is given by the Shakura-Sunyaev (1973) disk description, when the inner boundary condition is neglected
\begin{equation}
F = {3 G M \dot M \over r^3}
\end{equation}
and this is the same as Eq.~\ref{eq:teff}.

Eq.~\ref{eq:dynamics} does not include whole complexity of the dust motion under the influence of the gravity
and radiation pressure (see e.g. Netzer \& Marziani 2010). In particular, it applies only when $z << r$,  it neglects the radiation pressure due to the central source and it also neglects the change of the dust opacity with the temperature of the disk emission. However, the advantage of this oversimplification is that this equation can be solved analytically, and the solution is particularly simple.

We obtain the following expressions for the height of the dust particle above the disk and its vertical velocity as a function of time $t$
\begin{equation}
z(t) = z_*(1 - \cos(\Omega_K t)),
\end{equation}
\begin{equation}
v(t) = z_* \Omega_K \sin(\Omega_K t), 
\end{equation}
where $\Omega_K$ is the local Keplerian velocity and the constant $z_*$ is given by
\begin{equation}
z_* = {3 \dot M \sigma_{dust} \over 8 \pi m c}.
\label{eq:zstar}
\end{equation}

Here we additionally neglected the disk thickness $H_{disk}$ although the analytic solution exists for $H_{disk} > 0$; in that case the amplitude of the oscillation is smaller but the mean values remain the same.

The dust grains/clouds in this approximation perform symmetric oscillatory motion, up and down, in the vertical direction above the mid-plane. They maximally reach the hight of $2z_*$, independently from the radius $r$ where they are located.

Since the gravitational attraction and the radiation flux scale both in the same way with the disk radius, $r$, the height reached by the dust is independent from the radius in this approximation. This means that the highest relative angular coverage, $z/r$, is reached at the innermost radius of the BLR in this approximation.

The maximum velocity is reached at $z = z_*$, and it is equal to $z_* \Omega_K$. Therefore, all the interesting
parameters are set by the ratio of the $z_*/r_{\rm BLR}$, and this ratio should be smaller than 1 for our analytical solution to hold.

Independently, the material rotates around the symmetry axis with the Keplerian speed, as the rest of the accretion disk. The vertical velocity in this picture is never higher than the Keplerian speed, and this ratio again is the highest at the inner radius of the disk.

We further assume that the plasma emissivity is the same for all clouds, and we further simplify the situation by assuming that the whole moving dusty plasma consists of the randomly distributed clouds, moving with Keplerian speed around the symmetry axis, and with a Gaussian distribution of the vertical velocities centered at zero. The number of clouds scales as an arbitrary power law with the radius: the simple model does not predict the outflow/inflow rate. To make things more realistic, we included the moon-type cloud visibility effects (only the cloud hemisphere facing the central region contributes to the spectrum). The total number of clouds was set to 50 000.

\begin{figure}
\label{fig:LBQS_new}
\begin{center}
\includegraphics*[width=10cm,angle=0]{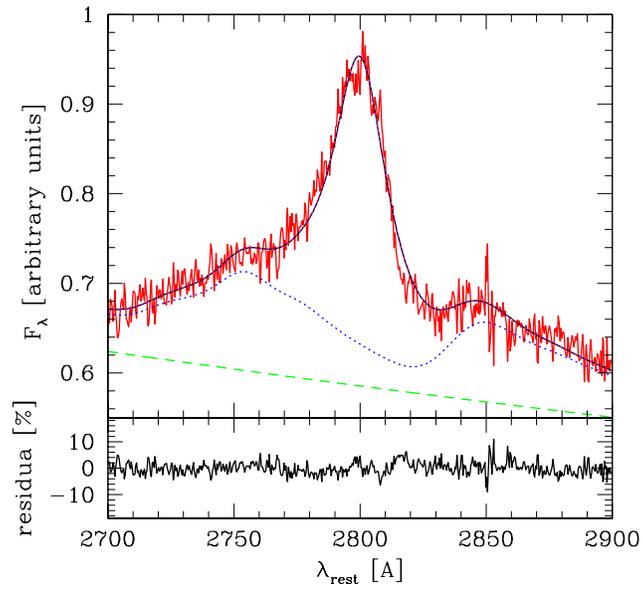}
\end{center}
\caption{The SALT spectrum of the quasar LBQS 2113-4538 in the 2700 - 2900 \AA~ rest frame (continuous line), and its decomposition into Mg II line modeled by a Lorentzian, Fe II pseudo-continuum from the set by Bruhweiler \& Verner (2008), and the powel law.}
\end{figure}

We use this setup to explain the simplest shapes of the lines observed in Narrow Line Seyfert 1 galaxies and in their high mass analogs - type A quasars (Sulentic et al. 2007). It is well known that LIL in these objects are well described with a single Lorentzian shape (e.g. Laor et al. 1997, Veron-Cetty et al. 2001, Sulentic et al. 2002, Zamfir at el. 2010, Shapovalova et al. 2012). We confirm this property with high accuracy spectrum of a quasar LBQS 2113-4538  (redshift 0.956) obtained with the 11-m Southern African Large Telescope (SALT). The source was observed on three occasions between May 15 and November 18, 2012. The spectra were discussed in Hryniewicz et al. (2014). Here we reanalyzed these spectra using the ESO spectral standard LTT1020 while another star (TYC 8422-788-1, located in the slit) was used by Hryniewicz et al. (2014) for recalibration. The shape did not change significantly, and the mean spectrum in the 2700 - 2900 \AA ~ spectral range again is well described by a power law, one of the Fe II pseudo-continua from Bruhweiler \& Verner (2008), and the Mg II line. The line shape was again well represented as a single kinematic component of a Lorentzian shape, the same for the two doublet components (see Fig.~4). The new value of the Mg II equivalent width ($EW$) is 21.28 \AA~ instead of 21.33 \AA~ in Hryniewicz et al. (2014) which shows that the calibration details do not influence the results significantly.

\begin{figure}
\label{fig:Petrogalli}
\begin{center}
\includegraphics*[width=10cm,angle=0]{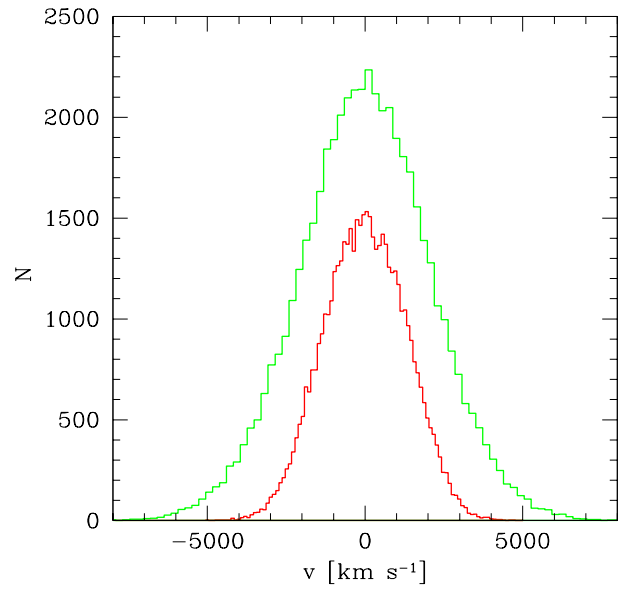}
\end{center}
\caption{The FRADO model can roughly reproduce the Lorentzian shape of the Mg II if the 
maximum vertical velocity is very high and the power law index in the radial distribution of the number of clouds is steep; 
two examples: $z_*/r_{\rm BLR} = 0.5$ (red line), and $z_*/r_{\rm BLR} = 1.0$ (green line)}
\end{figure}

This simple Lorentzian shape was reproduced by the model, when we used very large values of  $z_*/r_{\rm BLR} = 0.5$, i.e. of the vertical extension and the vertical velocities (see Fig.~4). Such a ratio is not unrealistic in the case of high accretion rate and large black hole mass objects. To see that we expressed the ratio  $z_*/r_{\rm BLR} $ in terms of the optical depth for dust and for electron scattering, instead of using the dust cross section and the cloud mass present in Eqs.~6 and 10
\begin{equation}
{z_* \over R_{\rm BLR}} = 0.56 \left({M \over 10^8 M_{\odot}}\right)^{1/3} \left({\dot M \over \dot M_{\rm Edd}}\right)^{2/3} {\tau_{\rm dust} \over 10 \tau_{\rm Th}},
\end{equation}
where the Eddington accretion rate is defined as $\dot M_{\rm Edd} = 4 \pi G M m_p/(\sigma_{T} \eta c)$, $\sigma_{T}$ is the Thomson cross section, $m_p$ is the proton mass, $\eta$ is the flow efficiency, here adopted to be 0.1, and the optical depth of the cloud, $\tau_{\rm dust}$,  is measured in the vertical direction. Since the  optical depth due to the dust can be a few times larger than the Thomson optical depth, the value $z_*/r_{\rm BLR}$ can be large for the flow close to the Eddington rate. It is also interesting to note that this ratio depends explicitly also on the black hole mass. Therefore, larger heights are characteristic for quasars, with black hole masses in the range from $10^8 M_{\odot}$ to a few times $10^9 M_{\odot}$, but they will be by a factor of a few smaller for nearby Seyfert 1 galaxies, with masses of order $10^6 - 10^7 M_{\odot}$. The ratio also explicitly depends on the dust opacity. Therefore, in the case of higher metallicity objects, with likely enhanced dust content, the ratio would be higher and the line profiles  would have more Lorentzian shape.

In order to obtain the Lorentzian shape we also had to assume that the radial distribution of clouds follows a power law with an index of high negative value. This last effect is rather consistent with the model expectations: most of the clouds concentrate around the inner radius of the BLR. However, large velocities are not fully consistent with the assumptions underlying the model. Lower extensions/velocities show significant imprint of the disk-line shape. While it is acceptable for lower Eddington ratio sources (e.g. Eracleous et al. 2009)  they should not be seen in high Eddington sources.  Our simple model is at least consistent with this trend since vertical velocities/extensions rise with $\dot M$ (see Eq.~\ref{eq:zstar}). Perhaps the global 3-D computations decrease this effect since the classical global wind solutions (e.g. Murray \& Chiang  1997) also do not show a strong disk-like line shapes.

Additional problem is that a single Lorentzian shape of Mg II and H$\beta$ line is characteristic only for a fraction of objects, with the highest
Eddington rates: type A quasars (Sulentic et al. 2002) and NLS1. Other quasars and Seyfert galaxies shows more complicated line shapes. In type B quasars there is an additional component of unknown origin (see e.g. Sulentic et al. 2002), not present in Fe II (Modzelewska et al. 2014). The line is then assymetric
so its connection with the disk-like motion implied by FRADO is unclear. Many years of monitoring of the H$\beta$ line in NGC 4151 seems to suggest
a possibility of a spiral structure or large scale accretion disk perturbations which are likely to explain the variability of the red and blue wing in the 
timescales of years (Shapovalova et al. 2010). Interesting Mg II shape variation in the radio-loud source 3C 454.3  correlated with radio burst (Le\' on-Tavares et al. 2013) points toward the role of the jet irradiation and/or the presence of additional emitting region. The same issue exist in both H$\beta$ and Mg II since the line shapes measured in the same source (possible for a narrow range of low/intermediate redshift AGN) in
general show considerable similarity (e.g. Marziani et al. 2013).

\subsection{Total mass of the material emitting LIL part of the BLR}

The FRADO model allows also to estimate the mass content of the line emitting region. The BLR starts at the inner radius. The outer edge is at the
inner radius of the dusty/molecular torus. However, most of the emission comes from the vicinity of the inner radius, as argued in Sect. 3.3. Therefore, we can assume that the radial extension of the dominating zone is also of order of $R_{\rm BLR}$, i.e. the surface covered is
$2 \pi R_{\rm BLR}^2$. The vertical extension is given by Eq.~10 so we have an estimate of the volume. The key parameter, however, in mass determination is the local density. We can estimate the upper limit for the density from the fact that the total optical depth in the vertical direction 
due to the dust cannot be larger than 1 since much larger optical depth would prevent the radiation pressure to act efficiently. With this assumption we obtain the following expression for the total mass
\begin{equation}
M_{\rm BLR} = 2 \pi R_{\rm BLR}^2 {  \sigma_{dust}\over m_p \sigma_{T}}
\end{equation}
and it is equal to $  8 \times 10^2 M_{\odot}$ for a $10^8 M_{\odot}$ black hole, and the dust opacity ten times higher than the electron scattering opacity. It is in agreement with the estimates of the BLR mass by Baldwin et al. (2003). However, the scaling of the BLR size implies that the mass of the BLR in the FRADO model scales linearly with the monochromatic flux, and can be a factor of ten higher for luminous quasars and two orders of
magnitude lower in less luminous nearby Seyfert galaxies.

\subsection{Disappearance of the BLR in low luminosity sources}

The presence of the dusty torus shields the central parts from the view, which explains the existence of type 2 sources. In the highly inclined sources the broad lines are visible only in the polarized spectra, scattered towards the observer. These sources are now known as HBLR (Hidden BLR). However, there are also sources which do not show broad lines even in the polarized spectra (known as naked or true Seyfert 2 galaxies). Some of them may be even more inclined than the HBLR objects, but in a number of low luminosity AGN broad lines seem to be absent. The model should explain that. In the case of the FRADO model, as in the other models of the BLR formation, the lines are connected in some way with the presence of the optically thick Keplerian disk. However, the inner parts of the disk itself seems to disappear with the drop of the luminosity, and they are replaced with some form of an optically thin flow. The optically thick disk is not present around the Sgr A*, as well as in quiescent state of the cataclysmic variables and galactic binaries. This may correspond to an ADAF formation, or a more complex scenario of the accretion disk evaporation. 

The condition that the BLR disappears when there is no underlying optically thick accretion disk is common to all mechanisms listed in Sect.~\ref{sect:intro}, including the FRADO model discussed in this paper. Such tests thus do not differentiate between the BLR formation mechanisms but nevertheless could falsify all the proposed mechanisms.  The studies also tests our detailed understanding of the disk transitions from the optically thick to the optically thin state and back.

The number of the low luminosity sources still showing the broad line component is not high. The most recent  paper along these lines (Balmaverde \& Capetti 2014) showed that the simple ADAF does not well reproduce well the transition, but the disk evaporation model with the magnetic field effects included taken from Czerny et al. (2004) well delineates the presently known objects with and without the broad lines.

Thus the lack of the underlying disk prevents the existence of the BLR but the existence of the disk may not be sufficient. In the magnetic wind models the torque of the outflowing wind must be consistent  with energetic requirements, and such a condition forms an independent limit on the parameters of the sources with broad lines (Elitzur \& Shlosman 2006, Elitzur \& Ho 2009). In their recent paper Elitzur et al. (2014) point out that the material has to rise well above the disk in order to get irradiated. If we apply such a criterion to the FRADO model and assume that the covering factor of the failed wind has to be higher than certain minimum value, we obtain a necessary condition for the BLR existence expressed in term of the bolometric luminosity (i.e. $ \dot M$) by combing Eqs.~\ref{eq:zstar} and \ref{eq:RBLR1}
\begin{equation}
L_{bol,min} \propto M^{1/2}.
\end{equation}

The slope in this relation is somewhat different from the relation found by Elizur \& Ho (2009) from their magnetic wind model
\begin{equation}
L_{bol,min} \propto M^{2/3},
\end{equation} 
but the present data collected by Elitzur et al. (2014) does not allow yet to fix the slope in this relation from observations.

Clearly, more low luminosity AGN with broad lines are needed but the task is difficult due to the dominance of the host galaxy in these sources, including the circumnuclear starburst.

\section{Discussion}

The Failed Radiatively Accelerated Dust-driven Outflow (FRADO) model is an attractive new way to explain the basic mechanism of the formation of the BLR in active galaxies. It is very simple and explains the basic observational facts, like the scaling of the size of the BLR with the square root of the monochromatic luminosity, the presence of the material well above the disk,  and the absence of strong outflow signatures in the lines like H$\beta$ and Mg II.

However, the model is still in its infancy and has to be merged with the main stream of the disk wind research.  A lot of research study was done about the disk winds (Murray \& Chiang 1997, Proga et al. 2000, Proga \& Kallman 2004), some of the papers actually addressed the failed wind issue, although in the context of the line driven scenario (e.g. Risaliti \& Elvis 2010). Very interesting and important work on the dynamics of a single irradiated cloud was recently published by Proga et al. (2014), and Namekata et al. (2014). Gaskell \& Goosmann (2013) point out the importance of scattering in shaping the lines. The numerical models address so far only specific questions starting from a specific setup, and they are already very demanding from the computational point of view. However, when our understanding of key specific phenomena improves we will be able to formulate a global picture of the BLR.  A specific issue is also the dependence on the object metallicity. The FRADO model predicts some dependence of the BLR location and the line shape on metallicity and chemical composition of the dust. More silicates imply lower dust sublimation temperature, and smaller BLR radius.  Higher metallicity, likely leading to higher dust content, implies line shapes closer to a Lorentzian shape. Tests of models may be 
additionally complicated if the metallicity and luminosity are correlated. Such a correlation was suggested by Hamann \& Ferland (1993), and 
discussed more recently by Matsuoka et al. (2011) but other studies indicate no metallicity evolution in AGN (e.g. Juarez et al. 2009) so the issue is far from being set.

The current FRADO model in the next step has to be taken from 1-D  toy model to a full 3-D description of the matter movement and emissivity. It is clearly important since the clouds rising above the disk are finally illuminated by the intense radiation produced close to the central black hole.
The emission of the inner parts of the accretion disk roughly depends on the cosine of the inclination angle so the irradiation is negligible at disk grazing
angles but it increases as soon as material is considerably alleviated. It becomes equal to the local flux at the height $z$ approximately equal 
$3/(4 \eta)  R_{\rm Schw}$, where $\eta$ is the efficiency of the accretion flow. However, such a simple geometrical estimate is unlikely to be correct
for two reasons. First, the accretion disk shows significant self-shielding property for the Eddington ratios close to 1 or above (Czerny \& Elvis 1987, 
Wang et al. 2014).  Second, the opacity towards the nucleus at large inclinations is likely large, so the concept of the shielding by the innermost
outflow was extensively discussed in the literature (e.g. Gaskell 2009; Gallagher \& Everett 2007), and this effect would weaken the irradiation by the central source. Only a full
3-D picture allows to calculate such non-local effects.

A separate issue is a problem of self-gravity effects in the accretion disk. Within the FRADO model we neglect this issue. On the other hand, if the
luminosity of some quasars is considerably higher than the Eddington ratio, the self-gravity radius would be within the BLR or even closer in. This
would entirely change the disk structure (Collin \& Zahn 1999,  Wang et al. 2012) and would affect the expected line profiles.

The progress in modelling is thus strongly required, but the final progress cannot be reached without a new stream of observational data. Only in confrontation with the data the key parameters involved can be set, like the issue of the strength and the role of the magnetic field in comparison with the radiation pressure force.

The need for more data means more known and well studied AGN in a single epoch spectroscopic studies, more time-resolved spectral studies, including reverberation monitoring, more well covered broad band data to give the hand to better modelling of the incident continuum for the BLR.

The number of known and modeled AGN is already impressive, there are close to 100 000 objects in SDSS (Schneider et al. 2010, Shen et al. 2011). However, most of the current effort in reverberation studies still concentrates on the nearby AGN since the variability timescales, and the requested duration of the monitoring, is then shorter, like a fraction of a year (see e.g. Barth et al. 2013; Rafter et al. 2013, Du et al. 2014, and the references therein).  First very preliminary results, however, are available for a few selected more distant sources (Kaspi et al. 2007; Perna et al. 2014), and more is expected to come (e.g. Shen et al. 2014). The observational results about the continuum variability and the response of the BLR lines clearly show the phenomenon complexity. Gaskell (2011) as well as Ruan et al. (2014) discuss the possibility of the off-axis localized variability. Grier et al. (2013) demonstrate the complexity of the BLR motion seen from the velocity-delay maps, with inflow likely to be present, as previously seen in Fe II lines (Ferland et al. 2009). More such high-quality results are urgently needed.

GAIA mission will increase the number of known quasars by a factor of 5, up to 500 000. Only very low resolution spectroscopy will be performed, but perhaps it will still be possible to measure the line intensity well enough to allow the measurements of the time delays between the line and the continuum. Since the sky will be covered approximately 70 times during the 5 year mission, this may be enough for brighter quasars with higher spectral quality. The survey will also bring more AGN at the border between broad line objects and Low Luminosity AGN.

With JWST (James Webb Space Telescope, planned to be launched in 2018) we will observe numerous quasars in the IR, which will allow to follow the evolution of the nuclear activity, including the formation of the broad lines.

LSST is the planned photometric survey devoted to variability studies, and this ground-based intrument will surely bring excellent results for quasar monitoring.  The idea is to use two pairs of the photometric bands: one relatively free from strong emission lines and one with high contribution from one of the lines. Then the decomposition of the flux into line and continuum in the second band has to be made, which allows then to measure the time delay. The method works, as was already shown by Chelouche et at. (2014).

Finally, Athena satellite mission (planned to be launched in the late 2020s) will allow us to complement the broad band spectra with the X-ray part for many more sources than it was possible so far. It will also allow to measure the X-ray variability of distant AGN, and for the selected sources observed in X-rays it will be possible to determine the inclination based on the X-ray variance thus extending the study of a few objects shown in Fig.~3.

\section*{Acknowledgements}
The spectroscopic observations reported
in this paper were obtained with the Southern African Large Telescope
(SALT), proposals 2012-1-POL-008. JM, BC, FP,
MK, and A\' S acknowledge the support by the Foundation for Polish Science
through the Master/Mistrz program 3/2012. K.H. also thanks the Scientific Exchange
Programme (Sciex) NMSch for the opportunity of working at ISDC. Part of this work was 
supported by Polish grants Nr. 719/NSALT
/2010/0, UMO-2012/07/B/ST9/04425 and by Polish National Science Center project 2012/04/M/ST9/00780.


\end{document}